\documentclass[letterpaper, 10 pt, conference]{ieeeconf}  

\IEEEoverridecommandlockouts                              

\usepackage{graphicx} 
\usepackage{url}
\usepackage{cite}
\usepackage{algorithm}
\usepackage{algorithmicx}
\usepackage{algpseudocode}
\usepackage{amsmath,amssymb,amsfonts}
\usepackage{flushend}

\title{\LARGE \bf
Uncertainty Set Prediction of Aggregated
Wind Power Generation based on Bayesian LSTM and
Spatio-Temporal Analysis*
}

\author{Xiaopeng Li$^{1}$, Jiang Wu$^{1}$, Zhanbo Xu$^{1}$, Kun Liu$^{1}$, Jun Yu$^{1}$, Xiaohong Guan$^{1, 2}$ 
	\thanks{*This work was supported in part by the National Natural Science Foundation of China (61773308, 61903293).}
	\thanks{$^{1}$X. Li, J. Wu, Z. Xu, K. Liu, J. Yu and X. Guan  are with the MOE KLINNS Lab of Xi'an Jiaotong University, Xi'an, Shanxi, 710049, China (e-mail: lixiaopeng@stu.xjtu.edu.cn, jwu@sei.xjtu.edu.cn, zbxu@sei.xjtu.edu.cn, kliu@sei.xjtu.edu.cn, yu971212@stu.xjtu.edu.cn, xhguan@xjtu.edu.cn). }
	\thanks{$^{2}$X. Guan is also with the Center for Intelligent and Networked Systems (CFINS), Department of Automation, TNLIST, Tsinghua University, Beijing 100084, China.}%
}

\begin{document}
\UseRawInputEncoding
\maketitle
\thispagestyle{empty}
\pagestyle{empty}

\begin{abstract}

Aggregated stochastic characteristics of geographically distributed wind generation will provide valuable information for secured and economical system operation in electricity markets. This paper focuses on the uncertainty set prediction of the aggregated generation of geographically distributed wind farms. A Spatio-temporal model is proposed to learn the dynamic features from partial observation in near-surface wind fields of neighboring wind farms. We use Bayesian LSTM, a probabilistic prediction model, to obtain the uncertainty set of the generation in individual wind farms. Then, spatial correlation between different wind farms is presented to correct the output results. Numerical testing results based on the actual data with 6 wind farms in northwest China show that the uncertainty set of aggregated wind generation of distributed wind farms is less volatile than that of a single wind farm.

\end{abstract}

\section{INTRODUCTION}

Wind energy gets an important place among these resources due to its wide distribution, and wind energy source has seen increased use in recent years. In China, 71.7 GW of new wind capacity was installed in 2020\cite{url}.
However, because of the intermittence, randomness, large-scale wind energy causes great challenges to the safety and economic operation of the entire power system. Thus accurate forecasting is Essential.
Traditional wind power forecasting model including persistence model\cite{persistencemodeling}, auto-regression(AR) model\cite{arima}, KNN\cite{knn}, Support Vector Regression (SVR)\cite{svr}, Gaussian Process\cite{Gaussianprocess}, artificial neural networks (ANN)\cite{ANN}. All of these models are univariate models. In other words, these models rely on historical data, temporal, wind speed, from a single wind farm, focusing on temporal relations but neglecting spatial correlation. Since wind is a type of airflow propagating rapidly in a large-scale area, adjacent wind farms' power outputs are inherently correlated. Thus several spatio-temporal models are proposed. In\cite{fuzzymodel}, a fuzzy model is used which considering neighboring sites information. In \cite{9}, a model for the average wind speed two hours ahead is introduced by considering wind direction at different places. The core issue for spatial-temporal wind forecasting is to model the relationship between the historical wind power of surrounding farms and the forecasting values of the target farm.
Regardless of whether it is a temporal or a spatial-temporal model, the final output is fixed. However, in the dispatch of the power grid, more attention is paid to the boundary, within which the safest and most economical dispatching plan is made for the power grid.
Therefore, compared with traditional wind power forecasting, we would pay more attention to the upper and lower bound of the forecast, that is, the width of uncertainty, with which to provide useful boundary conditions for our subsequent large-scale grid dispatch and optimization problems.

Several approaches to quantify the uncertainty of prediction have recently developed. In \cite{10lee2012objective}, an approach ensembling forecasts based on (Numerical Weather Prediction)NWP models with different conditions are provided. In \cite{11wu2018probabilistic}, a method to use lower upper bound estimation(LUBE) combined with NN to obtain prediction interval is provided. And The Gaussian process (GP), which involves a full Bayesian learning algorithm, is also a widely used approach for probabilistic wind energy forecasting\cite{12}. But the computational cost is heavy. Thus constructing a probabilistic forecasting model is challenging.

In this paper, a Spatial-temporal model is proposed for the problem of uncertainty set prediction. Our model learns temporal correlation and spatial features in near-surface wind fields of neighboring wind farms. We use a model called Bayesian LSTM, in which weights are not being a fixed matrix, but distribution, and Bayes by Backprop (BBB) is introduced to train the model. Monte Carlo sampling is used to estimate the distribution of the prediction result. Thereby obtaining the upper and lower bounds of estimation. Simultaneously, considering the spatial correlation and in order to increase credibility, we further correct the result by analyzing the spatial features between neighboring wind farms. The test results proved our model achieved good performance, and we found that the uncertainty set of aggregated wind generation of distributed wind farms is less volatile than that of a single wind farm.

The remainder of the paper is organized as follows: Section II presents preliminaries; Section III shows our methodology, the details of our model will be presented. And Section IV describes our experiment and result. Section V concludes the paper.

\section{PRELIMINARIES}

\subsection{Notation}
Suppose there are $N$ wind farms, giving a period of $k$ timestamps, for the whole $N$ wind farms, the power output series is denoted as $X = [X^{0},X^{1},...,X^{s},...,X^{N-1}]^{T} \in\mathcal{R}^{N\times k} $, where $X^{s}=[x_{0}^s,x_{1}^{s},...,x_{k}^{s}]\in\mathcal{R}^{1\times k
}$, represent the $s$-th wind farm wind power data during the period of $k$ time stamps. As for uncertain width prediction for different wind farm, we use $W$ to indicate uncertain width. $W$ with superscript $s$, written as $W^{s}=[w_{0}^{s},w_{1}^{s},...,w_{k-1}^{s}] \in \mathcal{R}^{1\times k}$, to represent $s$-th wind farm uncertain width prediction for $k$-steps.

\subsection{Problem statement}
Traditional wind power forecasting problem is related to time series analysis(TSA). This problem is described as predicting future wind power data $\hat{x}_{T+n}$, which is also named $n$-step ahead prediction. Mathematically, the problem could be described as
$$
\hat{x}_{T+n}=f(X)=f([x_{1},x_{2},\cdots,x_{t},\cdots,x_{T}]) \eqno{(1)}
$$

Where $\hat{x}$ is the prediction at timestamp $T+n$, $f$ is the prediction method such as Holt-Winters, Auto Regressive Integrated Moving Average Model(ARIMA), Support Vector Regression(SVR). $X=[x_{1},x_{2},\cdots,x_{t},\cdots,x_{T}]$ is historical wind power data.

Nevertheless, in this paper, we predict the uncertain width rather than predicting the futures time series. The Problem could be described as
$$
\hat{W}_{T+n}=g(X)=g([x_{1},x_{2},\cdots,x_{t},\cdots,x_{T}])\eqno{(2)}
$$

Where $\hat{W}_{T+n}$ is Uncertain Width at timestamp $T+n$, $g$ is the method for prediction, and $X$, as already mentioned above, is the historical data. 
    
\section{METHODOLOGY}
In this section, we introduce our model for Uncertain Width Prediction. Roughly speaking, our model integrates the temporal part, which analyzes historical data's temporal relation, and the spatial part, which mining geographical features between wind farms. The general framework of the method is shown in Fig.1. The core part of the temporal analysis is the Bayesian LSTM, in which parameters are probability distribution, making it possible to measure confidence and uncertainty over prediction. In Fig.1, it is clear to see how the data flow in the model. Raw data $X$ collected from different wind farms pass through the temporal model, where we use Bayesian LS TM, a model that combined LSTM and Bayesian Neural Networks. The output $\hat{W}$ is the estimated uncertain width. During the spatial part, the previous result is corrected by exploiting the spatial features between wind farms.
\begin{figure}[thpb]
	\centering
	
	\includegraphics[width=0.48\textwidth]{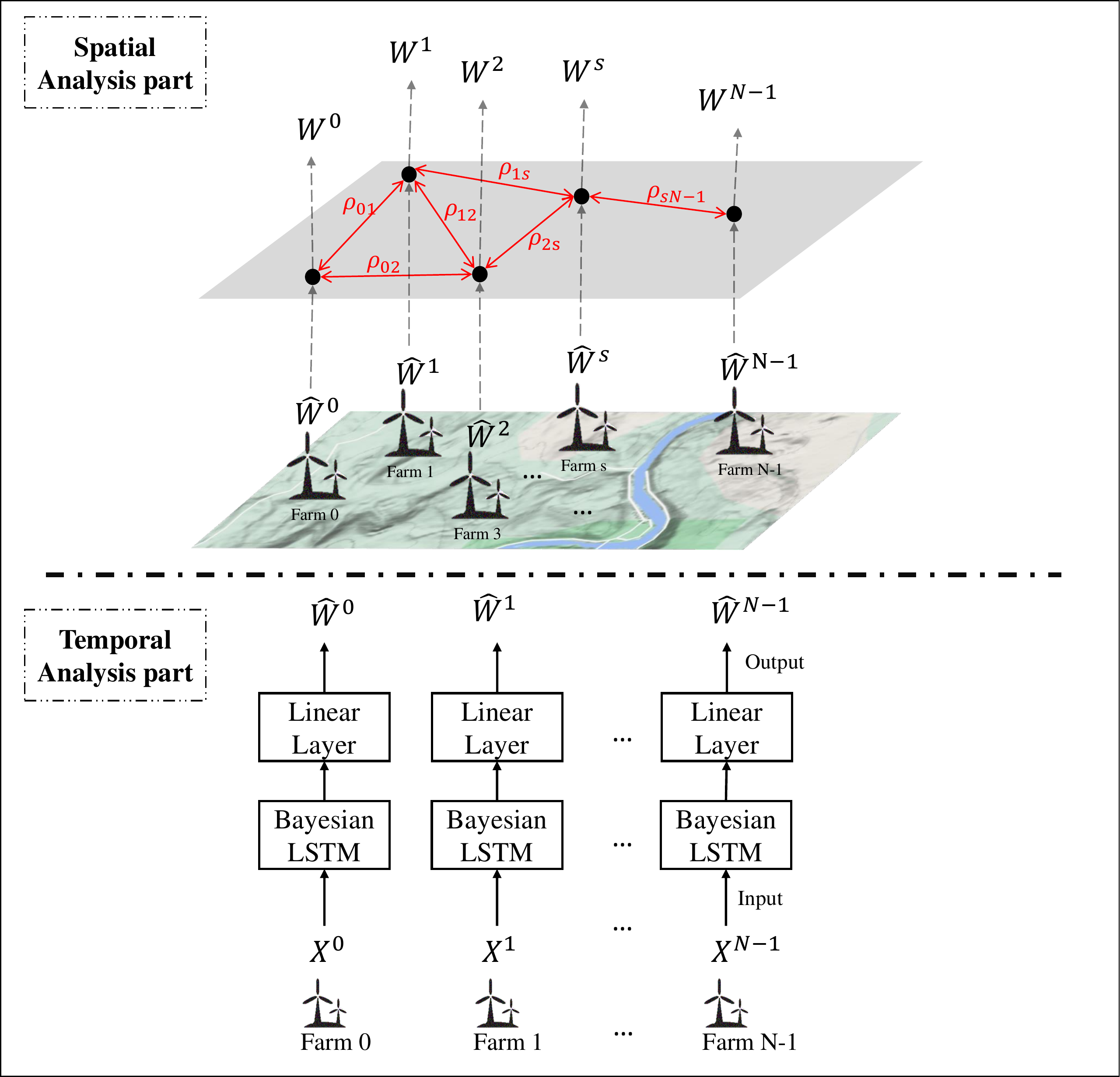}
	\caption{Structure of Our Proposed Model}
	\label{overallStructure}
\end{figure}

Thus, this section includes two categories: temporal analysis part-Bayesian LSTM, and spatial analysis.

\subsection{Temporal analysis- Bayesian LSTM}
Before we talk about Bayesian LSTM, it is necessary to take a brief review of LSTM and its Bayesian motivation.
\subsubsection{LSTM}
%
LSTM (Long Short-Term Memory) network is a special RNN (Recurrent Neural Networks) model. Compared with the general RNN model, its unique structure design avoids the long-term dependence problem.
The essential information of LSTM is the cell state, which is represented by $C_{t} $. The current LSTM receives the cell state from the previous moment and works with the signal input received by the current LSTM to produce the cell state of the current LSTM. 

LSTM uses a specially designed "gate" to clear or add information to the cell state. Gate is a way of letting information through selectively. LSTM mainly consists of three gate structures, forget gate, input gate, and output gate. 
These three gates are used to control the retention and transmission of LSTM information, which is ultimately reflected into the cell state $C_{t}$  and output signals $h_{t}$.
The forget gate consists of a sigmoid layer and a bitwise multiplication operation. It is used to forget unwanted information about the cell state. 
Contrary to forget gate, input gate would decide which of the newly input information $x_{t}$ will be retained. It is multiplied bit by bit with the tanh neural network layer results and then adds the result of forget gate to get the cell state's update values $C_{t}$.
In conjunction with the tanh function and the bit multiplication operation, the output gate transmits the cell state and input signals to the output $O_{t}$.

Following equations (3) formulate this process.

$$
\left\{ 
\begin{array}{l}

f(t)=\sigma(W_{f}\cdot[h_{t-1},x_{t}]+b_{f})\\

i(t)=\sigma(W_{i}\cdot[h_{t-1},x_{t}]+b_{i})\\
\tilde{C_{t}}=tanh(W_{c}\cdot[h_{t-1},x_{t}]+b_{c})\\
C_{t} = f_{t}\times C_{t-1}+i_{t} \times \tilde{C_{t}}\\
h_{t}=O_{t} \times tanh(C_{t})\\
O_{t} = \sigma(W_{0}\cdot[h_{t-1},x_{t}]+b_{o})

\end{array}
\right. 
 \eqno{(3)}
$$

\subsubsection{Bayes by Backprop}
We know that weights are deterministic for ANN (Artificial Neural Network), but for BNN (Bayesian Neural Networks), it samples weights from a probability distribution and then optimizes the distribution parameters.

\begin{figure}[thpb]
	\centering
	
	\includegraphics[width=0.48\textwidth]{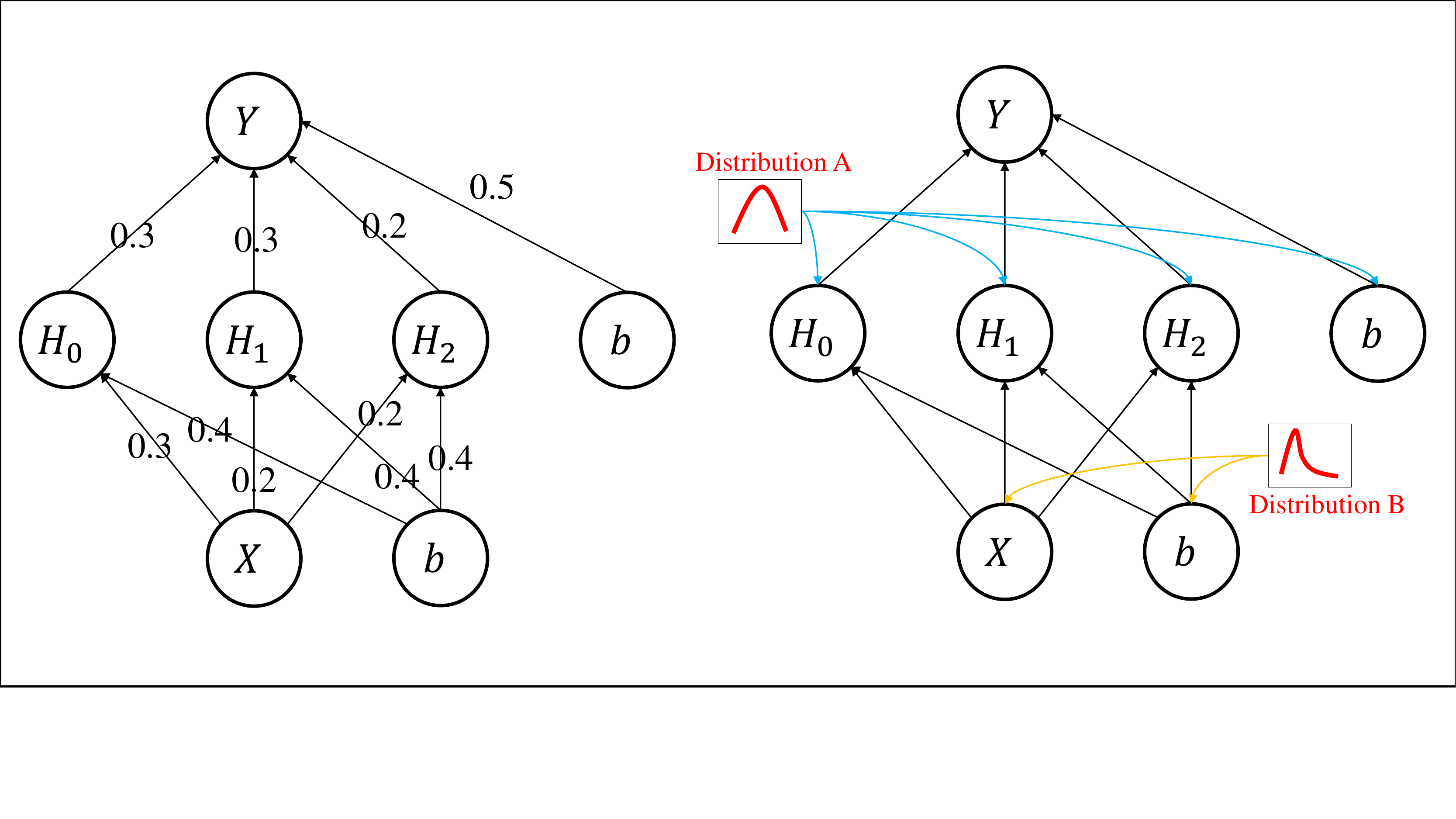}
	\caption{Left: ANN(Artificial Neural Network)weight with a fixed value. Right: BNN(Bayes Neural Network), weight is assigned to a distribution.}
	\label{Weights distribution}
\end{figure}

So, what we should care about is how to get the distribution of all the weights in the model. Charles et al. proposed Bayes by Backprop\cite{weightuncertainty}, this algorithm is for learning a probability distribution on the weights of a neural network. In the following paragraphs, details will be represented.

Bayesian inference for neural networks calculates the posterior distribution of the weights given the training data, $P(w|\mathcal{D})$, this distribution predict unseen data by taking expectations: the predictive distribution of an unknown label $\hat{y}$ of a test data item $\hat{x}$, is given by $P(\hat{y}|\hat{x})=E_{P(w|\mathcal{D})}[P(\hat{y}|\hat{x},w)]$, so we need to know what the posterior distribution $P(w|\mathcal{D})$ is,  variational inference is introduced, operated by constructing a new distribution $q(w| \theta )$ , over the weights $w$ and parameterized by $\theta$, that approximates the true posterior $P(w | \mathcal{D})$ . More concretely, it finds the setting of the parameters $\theta$ which minimizes the Kullback-Leibler (KL) divergence between $q(w| \theta )$  to $P(w|\mathcal{D})$.
$$
\theta^{*}=\mathop{\arg\min}_{\theta}KL[q(w|\theta)||P(w|\mathcal{D})] \eqno{(4)}
$$
It is computationally prohibitive to calculate (4). Thus, an approximation method, which is to approximate cost function using sampled weights, is used.
$$
F(D,\theta) \approx \sum_{i=1}^{N}{q(w^{(i)}|\theta)}-logP(w^{(i)})-logP(\mathcal{D}|w^{i})
 \eqno{(5)}
$$
where $F$ denote the cost function, $w^{(i)}$ denotes the $i$-th Monte Carlo sample drawn from $q(w^{(i)}|\theta)$. After obtaining the cost function, we can derive $\theta$ to get the gradient and update the parameters.

\begin{figure}[!t]
	\begin{algorithm}[H]
		\caption{Bayes by Backprop}
		\label{alg1}

		\begin{algorithmic}
			\State Sample $\epsilon \sim \mathcal{N}(0,I)$, $\epsilon \in \mathbf{R}^d$,where $I$ is the identity matrix
			\State Set Network parameters $\theta= \mu+\sigma\epsilon$
			\State Do forward propagation and back propagation as normal.
			\State Let $g$ be the gradient w.r.t.$\theta$ from backporation
			\State Let $g_{\theta}^{F}$,	$g_{\mu}^{F}$,$g_{\sigma}^{F}$ be the gradients of $F(D,\theta)$ with respect to $\theta$,$\mu$,$\sigma$ respectively.
			\State Update $\mu$ according to the gradient $g+g_{\theta}^{F}+g_{\mu}^{F}$
			\State Update $\sigma$ according to the gradient $(g+g_{\theta}^{F})\epsilon+g_{\sigma}^{F}$			 	
		\end{algorithmic}
	\end{algorithm}
\end{figure}

\subsubsection{Introduction of Bayesian LSTM}
As mentioned in the previous part, we already know how Bayes by Backprop (BBB) works. We will combine BBB with LSTM. We refer to the work done in\cite{13bayesianLSTM}. Bayesian Neural Network samples weights for a probability distribution and then optimizes these distribution parameters. Thus, it is possible to measure confidence and uncertainty over predictions, along with the prediction itself, which is exactly what we want to get.

In the Bayesian LSTM model, each weight parameter is a random variable with a Gaussian mixture prior, and the weight at each time step had the same distribution. 

So, to be more specific, comparing Bayesian LSTM model with traditional LSTM, as (3) shown, parameters $W_{f},b_{f},W_{i},b_{i},W_{c},b_{c},W_{o},b_{o}$ at each step are all assigned to with a Gaussian mixture prior distribution, rather than a fix value or matrix.Gaussian mixture prior distribution is shown as follows.
$$
P(w) =\pi\mathcal{N}(w\|0,\sigma_{1}^{2})+(1-\pi)\mathcal{N}(w|0,\sigma_{2}^{2}) \eqno{(6)}
$$
where $w$ is weight of the network. $\mathcal{N}(w\|0,\sigma^{2})$ is the Gaussian density distribution evaluted $w$ with mean $\mu$ and variance $\sigma^{2}$, $\pi$ is a combination coefficient.  
A diagonal Gaussian distribution is used as the variational posterior distribution, which is often computationally tractable and numerically stable. Assuming that the network weights are uncorrelated. The loss function of the proposed model consisted of two components. The first part is the Mean Squared Error (MSE) loss calculated by the predicted and the observed wind power values, which is the most commonly used loss for predicting continuous values. The second part is the Kullback-Leibler (KL) divergence between the posterior and the prior distribution used for penalizing model overfitting. Bayes by Backprop (BBB) is adopted to update the parameters. While minimizing the loss, The model is training via shuffled mini-batches.
\begin{figure}[!t]
	\begin{algorithm}[H]
		\caption{Bayesian LSTM Model Training via Bayes by Backprop}
		\label{alg1}
		\begin{algorithmic}
			\Require $\mathcal{D}=(X,y)$,epoch size $E$, batch size $B$, and learning rate $\alpha$
			\For{$epoch=0,\cdots,E-1$}
			\Repeat\\
			\quad\quad1. Sample mini batch of size $B$ from training data {D}\\
	 \quad\quad2. Sample $\epsilon \sim \mathcal{N}(0,I)$, where $I$ is the identity matrix\\
	 \quad\quad3. Set network parameters $\theta = \mu+\sigma\epsilon$, where $\mu$ and $\sigma$ are the mean and standard deviation,respectively.\\
	 \quad\quad4. Computer the gradients of MSE loss with respect \\\quad\quad to $\theta$ using normal back-propagation:  $g_{\theta}^{L}$.\\
	 \quad\quad5. Computer the gradients of $F(\mu,\sigma,\theta)=log\mathcal{N}(\mu,\sigma^{2})$ - $logp(\theta)$ with respect to $\mu,\sigma,\theta:g^{F}_{\mu}, g^{F}_{\sigma}, g^{F}_{\theta}$, where $p(\theta)$ is the Gaussian mixture prior. $\mathcal{N}(\mu,\sigma^{2})$ is posterior  distribution, typically taken to be Gaussian, with mean $\mu$ and standard deviation $\sigma$.
	 \\\quad\quad6. Update $\mu=\mu-\alpha\frac{g^{L}_{\theta}+g^{F}_{\theta}+g^{F}_{\mu}}{B}$\\
	 \\\quad\quad7.Update $\sigma=\sigma-\alpha\frac{g^{L}_{\theta}\epsilon+g^{F}_{\theta}\epsilon+g^{F}_{\sigma}}{B}$\\
	 
			\Until{all the mini-batches are sampled}
			\EndFor
			\State \Return Fitted network model $f_{\theta}$
		\end{algorithmic}
	\end{algorithm}
\end{figure}
Since Bayesian LSTM has been obtained, transforming the distribution of the parameters into the uncertainty analysis of the result is still a problem. Thus we introduce the Monte Carlo method by sampling the prediction and calculate their expectation $\hat{\mu}$ and standard deviation$\hat{\sigma}$, When the sampling times is large enough, we could regard them as $\mu$ and $\sigma$ in the normal distribution $\mathcal{N}(\mu,\sigma^{2})$.

\subsection{Spatial Analysis}
From the previous part, we get the predicted result, but very rough. In this part, we want to fuse the spatial features between each wind farm and its neighbors to made corrections to the result. 

This section includes two steps. The first part is to establish a graph $G=\left( V, E \right)$, where $V$ is the node of the graph, representing each wind farm, and $E$ is the edge of the graph, representing the correlation between wind farms. After establishing the graph, the second step is to dig out the relationship between each wind farm and its neighbors to correct the uncertain width. In reality, we know that the operating conditions of wind farms are closely related to the environment, so wind farms in the same are spatially related. Therefore, the size of the uncertain width of a wind farm's output is also influenced by others. Especially in extreme situations, when a wind farm's output power fluctuates drastically, the farms adjacent to it will also produce drastic fluctuations, making uncertain width changes. That is why we want to compensate by using spatial correlation to get a more accurate uncertain width.

\subsubsection{Graph Construction}

The first step is to establish a graph $G=\left( V, E \right)$ where $V$ is the set of wind farms, and then we calculate the Pearson correlation coefficient between each node. In statistics, the Pearson correlation coefficient is used to measure the correlation between two variables $X$ and $Y$, and its value is between -1 and 1. The closer to 1, the stronger a positive correlation between the two variables. We set the threshold value as $\theta$ = 0.7. When the value of Pearson's correlation coefficient is greater or equal to 0.7, these two nodes are connected.

\subsubsection{Uncertain Width correction}

After the graph is established in the first step, we make corrections to the uncertain width output by the Baysian LSTM. We think this correction is "lazy" because if wind farms operate in a stable condition in most circumstances, only when an extreme situation has occurred, such as heavy rain, storm, Blizzard, and other extreme weather conditions. When these situations happen, the uncertainty set or output of certain units will become unstable. Therefore, at this time, due to the lack of foresight in forecasting, compensation is necessary. Here we assume that only When the uncertain width difference between two wind farms with edge connection is greater than $\beta$, we think that compensation should be made. Thus for wind farm $i$ with its uncertain width $W_{i}$, if its Adjacent node uncertain width $W_{j}$, and $|W_{j}-W_{i}|> \beta$, 
then $W_{i}$=$  W_{i}+ \alpha \cdot (W_{j}-W_{i})$, here we choose 0.5 because We only make slight corrections without changing the order of magnitude, mean to increase the confidence level. 
\section{Experiment}
Experiments are conducted on real-world dataset to evaluate the performance of our proposed model. In this section, we introduce the dataset and baseline algorithms used along with the experimental setup and the evaluation.

\subsection{Dataset}

We use the wind power generation dataset collected from northwest China. This dataset records wind power generation of 6 wind farms with a 15 minutes sampling rate from Jan 1, 2018, to Mar. 26, 2019. Each wind farm file includes 43164 pieces of data.Fig.4 shows the location of the six wind farms on the map.
And, in order to verify our model, we divide the data into two sets. For each wind farm, we take data in 2018 as the training set and data in 2019 data as the test set.
And for each set, we split them with one-step forward data. This means we use the sliding window method to get feature $X$ and the next value as label $y$.We set the window size to 12. 

\begin{figure}[thpb]
	\centering

	\includegraphics[width=0.48\textwidth]{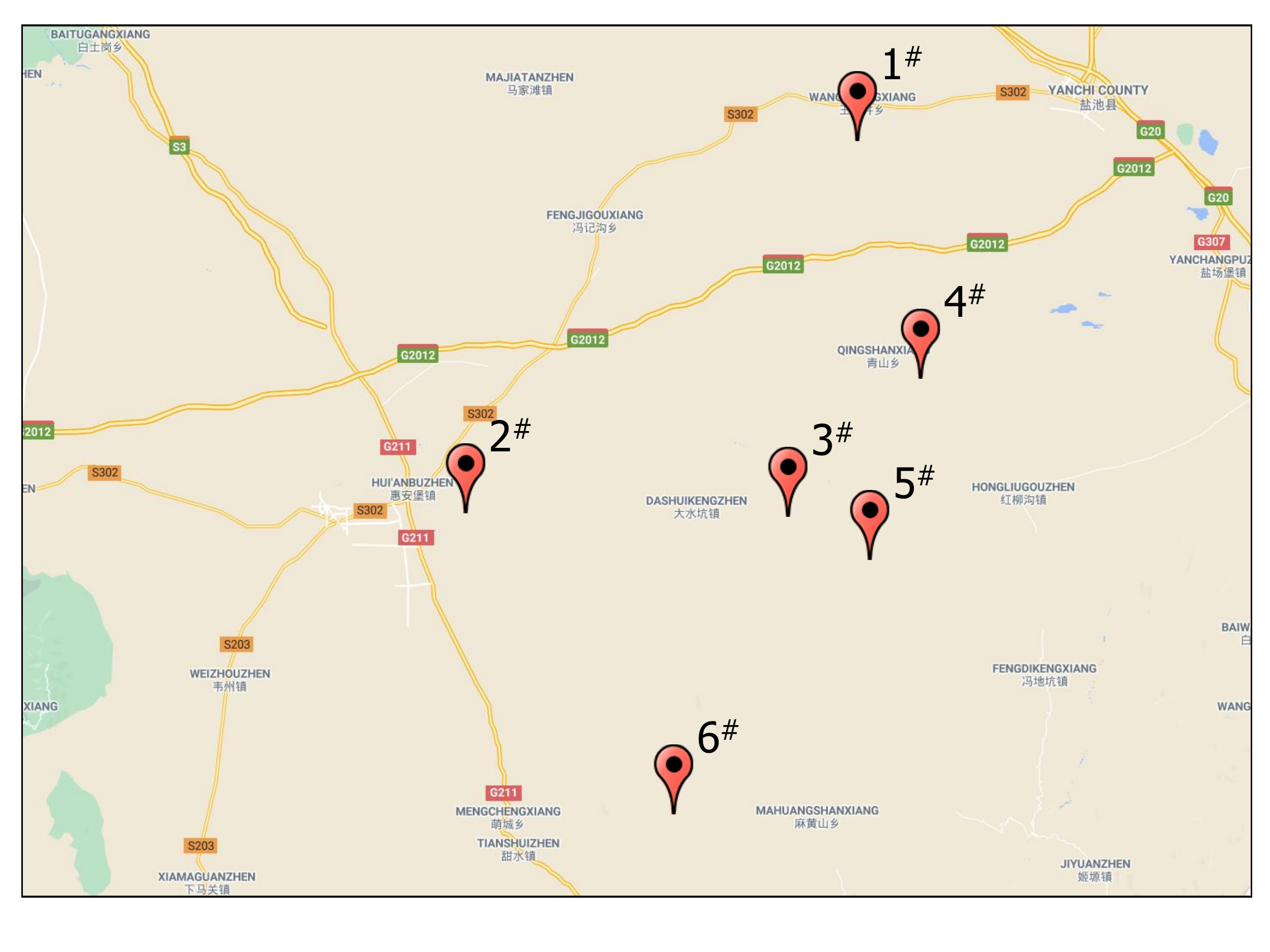}
	\caption{Location of wind Farms}
	\label{figurelabel}
\end{figure}
\subsection{Experiment Setting}

\begin{itemize}
\item Parameters setting in Temporal model: In this paper, we choose Baysian LSTM with 1 layer, hidden size equals 64, followed by a linear layer, remapping 64-dimensional data to 1-dimensional. As for prior distribution, we set $\pi$ and $\sigma_{1}$ as 1.

\item Parameters setting in Training: During training, we choose epoch number as 30, batch size as 800, MSE as loss function, Adam as an optimizer, learning rate as 0.001. And we complete it with GPU acceleration on a multi-GPU computer system with two NVIDIA GTX1080 graphics cards and a 2.10 GHz CPU.		
\end{itemize}

\subsection{Results}

This part will show and analyze the result from two aspects. The first is the result of uncertain width produced by the Temporal model-Bayesian LSTM. The second is the spatial correction result.
\subsubsection{Result of Bayesian LSTM}

After the training is completed, we send the prepared test data to the model for an experiment. For the Bayesian model, we need to obtain the test results through Monte Carlo sampling. Through sampling $N$ times, the predicted value is the expectation of the output value $\hat{\mu}=\mathbf{E}(y)$, where $y$ is sampling result. Then we calculate the standard deviation $\hat{\sigma}^{2}=\mathbf{D}(y)$ to get the upper and lower bounds of the prediction.
We assume that the data obeys the Gaussian distribution, then the 90\% confidence interval, the 95\% confidence interval, the 99\% confidence interval are calculated respectively as $I_{90\%} = \hat{\mu}\pm1.64\times \hat{\sigma}$,  $I_{95\%} = \hat{\mu}\pm1.96\times \hat{\sigma}$,  $I_{99\%} = \hat{\mu}\pm2.56\times \hat{\sigma}$. The uncertain width $W$ is the difference between the upper and lower bounds. 

Table \ref{Bayesian net out uncertain width} shows the uncertain width with 10 times sampling of Bayesian LSTM. And in order to further clarify the impact of Monte Carlo sampling times on the results, we take the sampling times as 10, 50, and 100 times to conduct comparative experiments on different confidence intervals and RMSE. The experimental results are as Table \ref{different times} shows.
As the number of sampling increases, the value of RMSE becomes larger, while the width of the interval gradually becomes smaller. This is also reasonable because the fewer the sampling times, the higher the randomness. Only when the sampling times increase, the uncertain width can decrease until it stabilizes near a certain value.


%

\begin{table}[h]
	\caption{Uncertain width with 10 times sampling of Baysiesian LSTM}
	\label{Bayesian net out uncertain width}
	\begin{center}
		\setlength{\tabcolsep}{0.6mm}{
		\begin{tabular}{|c||c||c||c|}
			\hline
			\textbf{Wind Farms} &\textbf{90\% confidence}&\textbf{95\% confidence}&\textbf{99\% confidence}\\
			\hline
			1\# & 14.78&17.67&23.08\\
			\hline
			2\# & 13.60&16.25&21.22\\
			\hline
			3\# & 11.70&13.98&18.26\\
			\hline
			4\# & 11.84&14.15&18.49\\
			\hline
			5\# & 12.09&14.44&18.86\\
			\hline
			6\# & 10.73&12.85&16.75\\
			\hline

		\end{tabular}}
	\end{center}
\end{table}

\begin{table}[h]
	
	\caption{Uncertain width with different sampling times for 1\# Wind Farm}
	\label{different times}

	\begin{center}
			\setlength{\tabcolsep}{0.8mm}{
		\begin{tabular}{|c||c||c||c|}
			\hline
			\textbf{Wind Farms} &\textbf{10 sample times}&\textbf{50 sample times}&\textbf{100 sample times}\\
			\hline
			90\% confidence & 10.73&9.36& 8.14\\
			\hline
			95\% confidence & 12.85 &11.20&9.73\\
			\hline
			99\% confidence & 16.75&14.62&12.70\\
			\hline
			RMSE & 9.12&9.50&12.70\\
			\hline			
		\end{tabular}}
	\end{center}
\end{table}

\subsubsection{Uncertain width correction}
In this section, we introduce corrections. 

In the first step, we establish the graph $G=\left( V, E \right)$ based on the data we have, which contains 6 nodes. And then, we calculate the Pearson correlation coefficient between each of them. The heatmap and building graph $G$ are shown in Fig.4.      
\begin{figure}[thpb]
	\centering

	\includegraphics[width=0.48\textwidth]{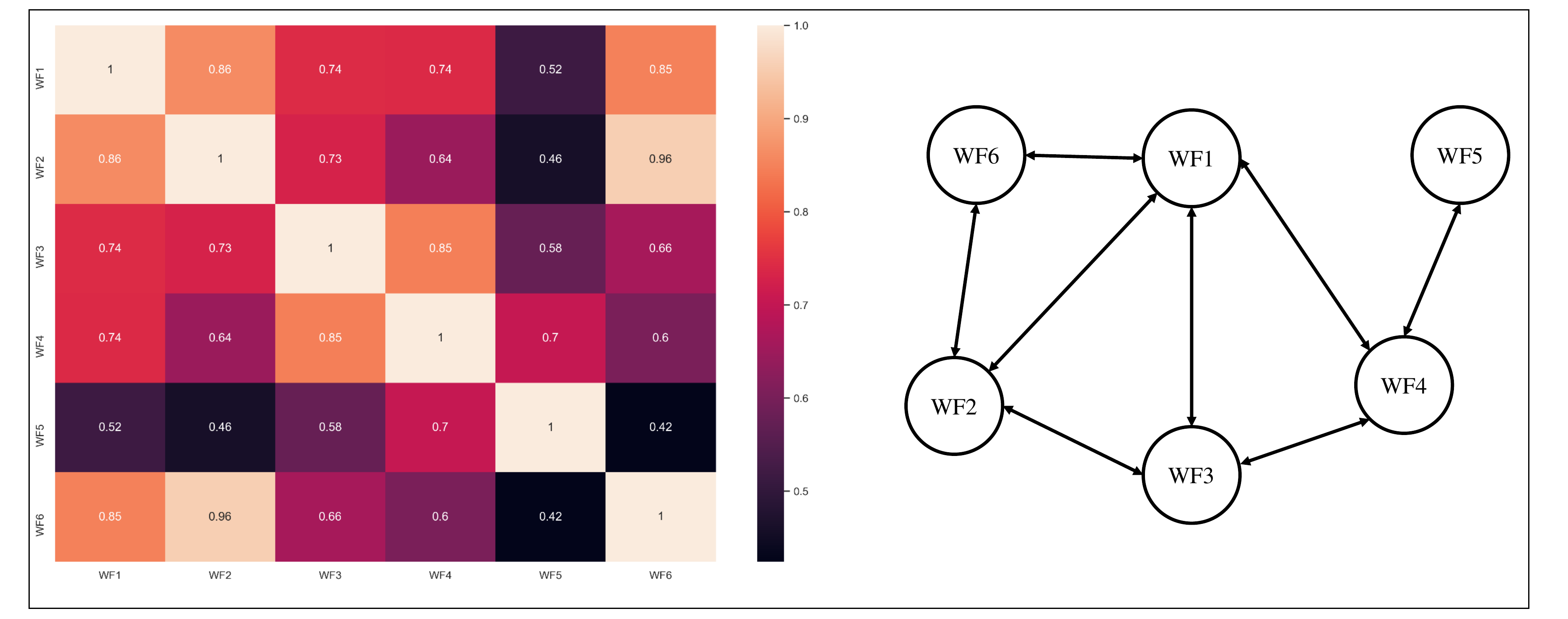}
	\caption{Left: The heat map of Pearson correlation coefficient. Right: Graph $G$ based on Pearson correlation coefficient.}
	\label{Heapmap}
\end{figure}
After constructing the graph, We correct the uncertain width according to the edge relationships. In the previous sections, we have already introduced how to correct. We set Threshold $\beta=20$ and $\alpha=0.5$. And the corrected result comparison is shown in Fig.5. Compared with the data before correction, for 1\#, we get a more accurate result. Before the correction, some anomalies are outside the confidence interval, while after correction, anomalies are all, with proved the necessity of spatial feature extraction.    
\begin{figure}[thpb]
	\centering
	\includegraphics[width=0.48\textwidth]{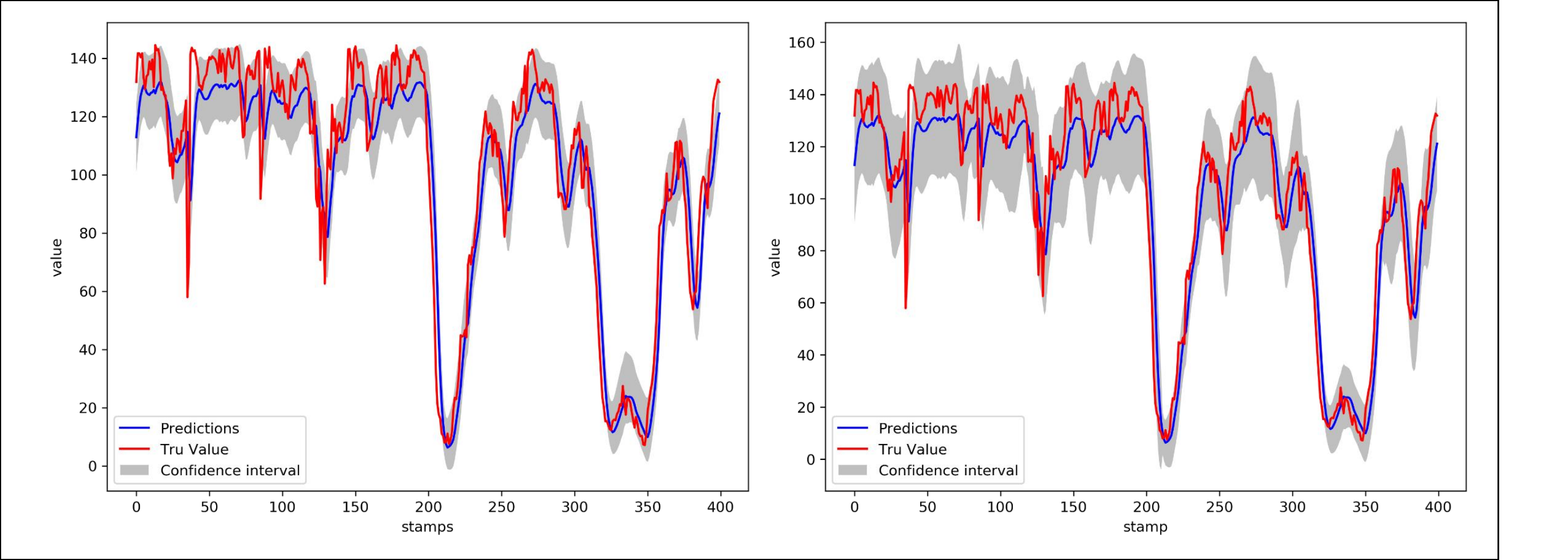}
	\caption{Left: 1\# windfarm Prediciton before correction. Right: 1\# windfarm Prediciton after correction.}
	\label{correctedresult}
\end{figure}

\begin{table}[h]
	\caption{99\% confidece interval After correction }
	\label{after correction}
	\begin{center}
		\begin{tabular}{|c||c|}
			\hline
			\textbf{Wind Farms} &\textbf{99\% confidece uncertain width after correction}\\
			
			\hline
			1\# & 30.12\\
			\hline
			2\# & 27.32\\
			\hline
			3\# & 24.58\\
			\hline
			4\# & 21.80\\
			\hline
			5\# & 19.76\\
			\hline
			6\# & 32.43\\
			\hline
		\end{tabular}
	\end{center}
\end{table}

\section{CONCLUSIONS}
In this article, We have established a spatiotemporal probability model that can predict the uncertain set of wind power data. Using Bayesian LSTM and building a spatial model by extracting correlation between wind farms, the uncertain set is predicted by Monte Carlo sampling. And we test our model on actual data with 6 wind farms in northwest China. And we will use the boundary value obtained by this method to dispatch the power grid accurately and economically in future works.

\addtolength{\textheight}{-12cm}   

\bibliographystyle{ieeetr}

\bibliography{ref}

\end{document}